\documentclass[12pt]{elsart}

\usepackage{amsmath}
\usepackage{amsfonts}
\usepackage{graphicx}

\begin{document}

\begin{frontmatter}

\title{Evidence of magic numbers in nuclei \\ by statistical indicators}

\author[rlr]{Ricardo L\'{o}pez-Ruiz} and
\ead{rilopez@unizar.es}
\author[jsr]{Jaime Sa\~{n}udo}
\ead{jsr@unex.es}

\address[rlr]{
DIIS and BIFI, Facultad de Ciencias, \\
Universidad de Zaragoza, E-50009 Zaragoza, Spain}

\address[jsr]{
Departamento de F\'isica, Facultad de Ciencias, \\
Universidad de Extremadura, E-06071 Badajoz, Spain, \\
and BIFI, Universidad de Zaragoza, E-50009 Zaragoza, Spain}


\begin{abstract}
The calculation of a statistical measure of complexity 
and the Fisher-Shannon information in nuclei is carried out in this work. 
We use the nuclear shell model in order to obtain the fractional occupation 
probabilities of nuclear orbitals.
The increasing of both magnitudes, the statistical
complexity and the Fisher-Shannon information, with the number of nucleons is observed.
The shell structure is reflected by the behavior of the statistical complexity. 
The magic numbers are revealed by the Fisher-Shannon information. 
\end{abstract}


\end{frontmatter}

\maketitle

In the last years, the study of the hierarchical organization of quantum systems by means
of information-theoretic indicators has taken a growing interest 
\cite{gadre1984,gadre1987,massen1998,chat2009}.
The application of these indicators to the electronic shell structure of atoms 
has received a special attention for systems running from one-electron 
atoms to many-electron ones as those corresponding to the periodic 
table \cite{coffey2003,dehesa2005,sanudo2008-,panos2005,borgoo2008,panos2007,romera2008}. 
Entropic products such as Fisher-Shannon information and
statistical complexity present two main characteristics when applied
to the former many-body systems. On one hand, they display an increasing trend 
with the atomic number $Z$ \cite{panos2005,borgoo2008}. 
On the other hand, they take extremal values on the closure of shells in noble 
gases \cite{panos2007,romera2008}.

Nucleus is another interesting quantum system that can be described by a shell model \cite{krane1988}.
In this picture, just as electrons in atoms,
nucleons in nuclei fill in the nuclear shells by following a determined hierarchy.
Hence, the fractional occupation probabilities of nucleons in the different nuclear orbitals
can capture the nuclear shell structure. This set of probabilities can be used
to evaluate the statistical quantifiers for nuclei as a function of the number of nucleons.
Similar calculations have been reported for the electronic atomic structure \cite{panos2009,sanudo2009}.  
In this work, by following this method, we undertake the calculation of statistical
complexity and Fisher-Shannon information for nuclei. 

The nuclear shell model is developed by choosing an intermediate three-dimensional potential, 
between an infinite well and the harmonic oscillator, in which nucleons evolve under the 
Schr\"odinger equation with an additional spin-orbit interaction \cite{krane1988}.
In this model, each nuclear shell is given by $(nlj)^w$, 
where $l$ denotes the orbital angular momentum ($l=0,1,2,\ldots$), 
$n$ counts the number of levels with that $l$ value,
$j$ can take the values $l+1/2$ and $l-1/2$
(for $l = 0$ only one value of $j$ is possible, $j=1/2$), 
and $w$ is the number of one-type of nucleons (protons or neutrons)
in the shell $(0\leq w\leq 2j+1)$. 

As an example, we explicitly give the shell configuration of a nucleus formed
by $Z=20$ protons or by $N=20$ neutrons. In both cases, it is obtained \cite{krane1988}:
 \begin{equation}
 \left\{\begin{array}{c}
 (Z=20) \\
 (N=20)
 \end{array} \right\}
 :  (1s1/2)^2(1p3/2)^4(1p1/2)^2(1d5/2)^6(2s1/2)^2(1d3/2)^4. 
 \end{equation}

When one-type of nucleons (protons or neutrons) in the nucleus is considered, 
a fractional occupation probability distribution of this type of nucleons 
in nuclear orbitals $\{p_k\}$, $k = 1,2,\ldots,\Pi$,  being $\Pi$ the 
number of shells for this type of nucleons, can be defined in the same way 
as it has been done for electronic calculations in the atom in \cite{panos2009,sanudo2009}. 
This normalized probability distribution $\{p_k\}$ $(\sum p_k=1)$ is easily found by dividing 
the superscripts $w$ by the total of the corresponding type of nucleons ($Z$ or $N$). 
Then, from this probability distribution, the different statistical magnitudes 
(Shannon entropy, disequilibrium, statistical complexity and
Fisher-Shannon entropy) can be obtained.

In this work, we undertake the calculation of a statistical measure of complexity, $C$, 
the so-called LMC complexity \cite{lopez1995,lopez2002}, that has been recently computed 
on some few-body quantum systems \cite{sanudo2008-,sen2008,sanudo2008+,howard2008} and 
also on some many-body quantum systems \cite{panos2005,panos2009,sanudo2009}. 
It is defined as 
\begin{equation}
C = H\cdot D\;,
\end{equation}
where $H$, that represents the information content of the system, 
is in this case the simple exponential Shannon entropy \cite{lopez2002,dembo1991},
\begin{equation}
H = e^{S}\;,
\end{equation}
$S$ being the Shannon information entropy. $D$ gives an idea of how much 
concentrated is the probability distribution of the system. 
The discrete versions of expressions $S$ and $D$ used in our calculations
are given by \cite{lopez1995}
\begin{eqnarray}
\hspace{4.3cm} S & = & -\sum_{k=1}^{\Pi}p_k\log p_k \; , \\
\hspace{4.3cm} D & = & \;\;\sum_{k=1}^{\Pi}(p_k-1/\Pi)^2 \; .
\end{eqnarray}

The Fisher-Shannon information, $P$, that has also been applied in atomic 
systems \cite{sanudo2008-,sen2008,sanudo2008+,howard2008,vignat2003,romera2004,szabo2008}, 
is given by
\begin{equation}
P = J\cdot I \; ,
\label{eq-p}
\end{equation}
where $J$ is a version of the power Shannon entropy \cite{dembo1991}
\begin{equation}
J = {1\over 2\pi e}\; e^{2S/3}\;,
\end{equation}
whereas $I$ is the so-called Fisher information measure, 
that quantifies the narrowness of the probability density. 
We use the same discrete version of $I$ as in \cite{panos2009}
\begin{equation}
I = \sum_{k=1}^{\Pi} {(p_{k+1}-p_k)^2\over p_k}\;,
\end{equation}
where $p_{\Pi+1}=0$.

\begin{figure}[h]  
\centerline{\includegraphics[width=12cm]{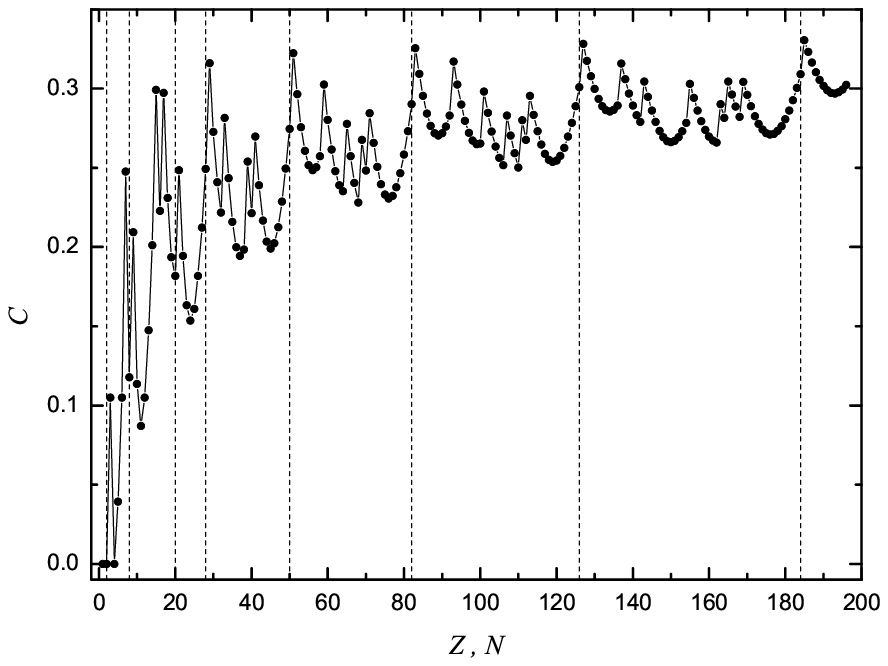}}  
\caption{Statistical complexity, $C$, vs. number of nucleons, $Z$ or $N$.
The dashed lines indicate the positions of magic numbers
$\{2,8,20,28,50,82,126,184\}$. For details, see the text.}  
\label{fig1}  
\end{figure}

The statistical complexity, $C$, of nuclei as a function of the number of nucleons, 
$Z$ or $N$, is given in Fig. \ref{fig1}. 
We can observe in this figure that this magnitude fluctuates around an increasing average value 
with $Z$ or $N$. This trend is also found for the electronic structure of atoms \cite{sanudo2009}, 
reinforcing the idea that in general complexity increases with the number of units forming a system.
However, the shell model supposes that the system encounters certain ordered rearrangements 
for some specific number of units (electrons or nucleons).
This shell-like structure is also unveiled by $C$ in this approach to nuclei.
In this case, the extremal values of $C$ are not taken just on the closed shells as 
happens in the noble gases positions for atoms, if not that they appear to be in the
positions one unit less than the closed shells.

\begin{figure}[h]  
\centerline{\includegraphics[width=12cm]{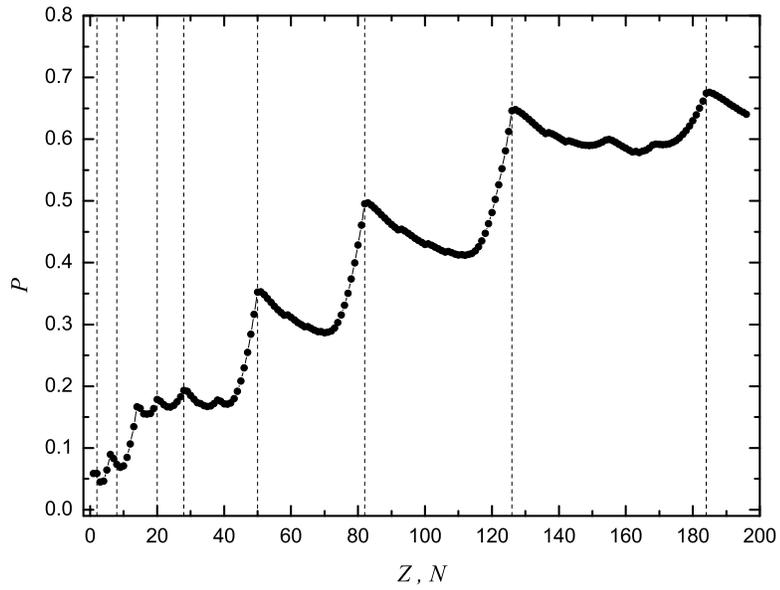}}     
\caption{Fisher-Shannon entropy, $P$, vs. the number of nucleons, $Z$ or $N$.
The dashed lines indicate the positions of magic numbers
$\{2,8,20,28,50,82,126,184\}$. For details, see the text.}  
\label{fig2}  
\end{figure}

The Fisher-Shannon entropy, $P$, of nuclei as a function of $Z$ or $N$ is given in Fig. \ref{fig2}. 
It presents an increasing trend with $Z$ or $N$. The spiky behavior of $C$ provoked by 
the nuclear shell structure becomes smoother for $P$, that presents
peaks (changes in the sign of the derivative) only at a few $Z$ or $N$, 
concretely at the numbers $2,6,14,20,28,50,82,126,184$. Strikingly, the sequence of
magic numbers is $\{2,8,20,28,50,82,126,184\}$ (represented as dashed vertical lines
in the figures). Only the peaks at $6$ and $14$ disagree with the sequence of magic numbers,
what could be justified by saying that statistical indicators work better for large numbers.
But in this case, it should be observed that the carbon nucleus, $C_{Z=6}^{N=6}$, and the 
silicon nucleus, $Si_{Z=14}^{N=14}$, apart from their great importance in nature and industry, 
they are the stable isotopes with the highest abundance in the corresponding isotopic series, 
$98.9\%$ and $92.2\%$, respectively.

Then, as it was suggested in a previous article \cite{sanudo2009},
the method that uses the fractional occupation probabilities can be applied 
to calculate statistical indicators, such as $C$ and $P$, in many particle systems.
Here, the behavior of the statistical complexity and the Fisher-Shannon
information with the number of nucleons in nuclei has been reported.
We have found the increasing trend of these magnitudes with $Z$ or $N$,
and the reflect of the shell structure in the spiky behavior of their plots.
It is worth to note that the relevant peaks in the Fisher-Shannon information
are revealed to be just the series of magic numbers in nuclei.
This fact indicates that statistical indicators are able to enlighten 
some structural aspects of quantum many-body systems.

\section*{Acknowledgements}
 The authors acknowledge some financial support from  the Spanish project
 DGICYT-FIS2006-12781-C02-01.

\newpage  

\newpage  

\end{document}